\documentclass[useAMS,usenatbib]{mn2e}
\input{psfig}                              %
\def\ie{\mbox{\'{\i}}}

\def\cicam{\mbox{CI Cam}}

\def\gro{\mbox{GRO J1655-40}}
\def\grs{\mbox{GRS 1915+105}}

\def\saxjdhdn{\mbox{SAX J1819.3-2525}}

\def\vqsqu{\mbox{V4641 \mbox{Sgr}}}

\def\cmmoinsdeux{\mbox{ cm}^{-2}}

\def\microns{\mbox{ } \mu \mbox{m}}

\def\kpc{\mbox{ kpc}}

\def\kms{\mbox{ km\,s}^{-1}}

\def\Msol{\mbox{ }M_{\odot}}

\def\keV{\mbox{ keV}}

\def\WMMUM{\mbox{ W m}^{-2}\mu \mbox{m}^{-1}}

\def\ergs{\mbox{ erg\,s}^{-1}}

\def\deg{^{\circ}}

\def\amin{^\prime}

\def\asec{^{\prime \prime}}
\def\asecp{{\rlap.}^{\prime \prime}}

\def\heu{^{h}}

\def\hmin{^{m}}

\def\hsecp{{\rlap.}^{s}}

\def\nh{N_{\rm H}}

\def\ltsima{\; \buildrel < \over \sim \;}
\def\simlt{\lower.5ex\hbox{\ltsima}}            
\def\gtsima{\; \buildrel > \over \sim \;}
\def\simgt{\lower.5ex\hbox{\gtsima}}            

\usepackage{amssymb}
\begin{document}

%
   \title[Observations of $\vqsqu$]
{Optical and near-infrared observations of the microquasar 
$\vqsqu$ 
during the 1999 September outburst
\thanks{Based on observations collected at the 
European Southern Observatory, La Silla, Chile, under proposal
numbers ESO N$\deg$ 63.H-0493 and 64.H-0382}}


   \author[S. Chaty et al.]
{S.~Chaty $^{1,2}$, P.A.~Charles $^3$, J.~Mart\'{\i} $^4$, 
I.F~Mirabel $^{2,5}$, 
L.F.~Rodr\'{\i}guez $^6$, \and T.~Shahbaz $^7$ \\ 
%
%
$^1$ {Department of Physics and Astronomy, 
The Open University, Walton Hall,
Milton Keynes, MK7 6AA, United Kingdom \thanks{chaty@cea.fr}} \\
$^2$ {Universit\'e Paris 7 and Service d'Astrophysique, 
CEA-Saclay, F-91191 Gif-sur-Yvette, Cedex, France} \\
$^3$ {Department of Physics and Astronomy, University of Southampton,
Southampton, Hampshire SO17 1BJ, United Kingdom} \\
$^4$ {Departamento de F\'{\i}sica, Escuela Polit\'ecnica Superior, 
        Universidad de Ja\'en, C/ Virgen de la Cabeza, 2, E-23071 Ja\'en, 
Spain} \\
$^5$ {Instituto de Astronom$\ie$a y F$\ie$sica del Espacio, Conicet, 
C.C. 67, Suc. 28 (1428),
        Buenos Aires, Argentina} \\
$^6$ {Instituto de Astronom\'{\i}a, Campus UNAM, Morelia, 
        Michoac\'an, 58190 M\'exico} \\
$^7$ {Instituto de Astrof$\ie$sica de Canarias, C/ V$\ie$a L\'actea, s/n, 
38205, La Laguna, Tenerife, Spain}
}
   \date{Received date; accepted date}
   \pubyear{2003} \volume{000} \pagerange{1} \twocolumn

   \maketitle \label{firstpage}

   \begin{abstract}
We present photometric and spectroscopic optical and near-infrared (NIR)
observations
taken during the outburst of the microquasar
$\vqsqu$ = $\saxjdhdn$ \citep{in'tzand:2000a} in September 1999.
We observed an increase in the J-K$_{\rm s}$ colour between 5 and 8 days 
after the outburst, which we interpret as likely evidence
for the presence of dust around the source.
We also observed an extraordinarily strong, broad and variable $H\alpha$ line,
with a velocity width of $4560 \kms$ suggesting the presence of a
high-velocity outflow component. 
We constrain the distance of the system between 3 and 8 kpc, 
locating it further away than previously derived from 
radio observations \citep{hjellming:2000}, but consistent with 
\cite{orosz:2001a}.
We then discuss the nature of this system, showing that
the companion star is either a B3-A2 main sequence star,
or a B3-A2 sub-giant crossing the Hertzsprung gap.
The system is therefore an Intermediate or 
High Mass X-ray Binary System (IMXB or HMXB).
The distance derived by these optical/NIR observations
implies that the jets observed by \cite{hjellming:2000}
would then exhibit apparent velocities of $\sim 10$ c.
We finally discuss the possibility of an interaction between
the jets and surroundings of the source, and also of this
source being a ``microblazar''.
%
%
   \end{abstract}

%

\section{Introduction}

The variable $\vqsqu$ attracted considerable attention after 
the detection of a giant optical outburst on 1999 September 15.7 UT, 
from V $\sim 14.0$ to 8.8 
\citep{stubbings:1999}.
Located in the direction of the galactic bulge
(galactic coordinates $l,b$ = $6.77\deg, -4.79\deg$),
$\vqsqu$ 
was initially confused with GM Sgr, and most of the references to this X-ray
source are reported under that name. 
After this confusion was clarified by \citet{williams:1999} and 
\citet{samus:1999},
the source was then designated $\vqsqu$ \citep{kazarovets:2000}.
The X-ray source XTE J1819-254 flared, from 1.6 to 12.2 Crab 
on 1999 September 14, as observed by {\it RXTE} in the 2-12 keV band, 
through a brief but dramatic eruption (at its peak it was the brightest
X-ray source in the sky), its position being coincident with
the optical transient \citep{smith:1999}.
Less than 10 hours later, the source was fainter than 50 mCrab.
It was also identified with the previously detected
faint X-ray transient, $\saxjdhdn$, 
discovered by {\it Beppo-SAX} on 1999 Feb. 20
(flux $0.012-0.3$ Crab, \citealt{in'tzand:1999}), 
and independently detected by {\it RXTE} (designated XTE J1819-254)
two days earlier \citep{markwardt:1999a}, with a flux
 between $3$ and $80$ mCrab in the 2-10 keV energy band.

Three other flares
followed, each lasting less than two hours in X-rays, some of the fastest
events ever seen. 
The observations by {\it RXTE} \citep{wijnands:2000}
allowed the observation of some strong flaring activity: 
fluctuations by factors of 4 and 500 on timescales of seconds and
minutes respectively. 
The spectrum was harder when the count rate 
was decreasing. During quiescent intervals, the
X-ray spectrum was much softer. 
No QPO was detected,
but some red noise at $<$ 30 Hz was present.
The observations by {\it Beppo-SAX} \citep{in'tzand:2000a} 
revealed a strong 
Fe-K emission line at $6.85 \pm 0.02 \keV$,
interpreted as fluorescence from a highly photo-ionized plasma.
A later analysis of the same data by \cite{miller:2002a} derived
an inclination of $i = 43\deg \pm 15\deg$.
No sign of eclipse or periodic signal due to binary orbit 
was detected during 19 hours, which was consistent with the 2.5 day period
found later.
The best fit for the column density was
$\nh = 0.05 \pm 0.02 \times 10^{22} \cmmoinsdeux$ \citep{in'tzand:2000a}.

The VLA radio telescope detected on Sept. 16.02 UT
a strong radio source (0.4 Jy) at 4.9 GHz, at the
position of the variable star
($\alpha$ = $18\heu 19\hmin 21\hsecp636$,
$\delta$ = $-25\deg 24\amin 25\asecp6$ J2000; \citealt{hjellming:1999a}).
The flux rapidly decreased on a timescale of hours, 
with an e-fold decay time of 0.6 days.
The source was resolved, with the presence of an elongation
extending $0.25 \asec$ between 0.6-1.2 days after the huge X-ray flare.
On Sept. 17.93, 22.00 and 24.1 UT, the elongation was at the same position
\citep{hjellming:1999b},
implying a proper motion of $0.5\asec$ / day, 
but this is strongly
dependent upon the time of the ejection. 
This allowed the source to be classified as a new microquasar
(for a review on jet sources see \citealt{mirabel:1999}).
An HI absorption measurement
towards the source implied a distance $d > 0.4 \kpc$ \citep{hjellming:2000},
and these authors proposed a likely distance of $0.5 \kpc$.

Goranskij (1978, 1990) reported a single short outburst of $\vqsqu$ in 1978,
recorded on Moscow photographic plates, reaching B = 12.4. He also
suggested the possible presence of a periodicity of 0.7365 days from
analysis of the quiescent data.
Another optical outburst occured in 1999 August 
\citep{watanabe:1999},
followed by a period of apparently increasing activity.
\cite{kato:1999} reported unusual optical activity 6 days prior to the 1999
Sept. giant optical and X-ray outburst, 
through a $\sim 1$ mag increase
combined with a modulation of 2.5 days, which they claimed to 
correspond to the orbital period.
More recently, during 2002 May, $\vqsqu$ was active again,
exhibiting in the optical chaotic 0.5 magnitude variations 
on a timescale of a few seconds.
In the radio domain, the source was also flaring on timescales of 
minutes to hours (M. Rupen, VSNET communication).
Details of this outburst are reported in 
{\it http://vsnet.kusastro.kyoto-u.ac.jp/vsnet/Xray/v4641sgr02.html}.

\cite{orosz:2001a} derived from ESO spectroscopic optical observations
of the source in quiescence 
a mass function of $2.74 \pm 0.12 \Msol$, which, combined
with their information on the inclination
($60 \leq i \leq 70\deg$, \citealt{orosz:2001a}),
makes $\vqsqu$ a black hole system
with a mass of the compact object in the range 
$8.73 \leq M_1 \leq 11.70 \Msol$. 
They also found a spectroscopic
period of $2.81678 \pm 0.00056$ days, and assuming an extinction
$E(B-V) = 0.32 \pm 0.10$, quoted a distance between $7.40$ and $12.31 \kpc$
(note that this is larger than previously derived by \citealt{orosz:2000}).

Through our on-going ESO Target of Opportunity (ToO) programme
 aimed at observing new X-ray flaring sources,
we quickly obtained near-infrared (NIR) and optical imaging 
and spectroscopic observations
of this new microquasar during the 1999 outburst and
we followed it during its decline from 1999 September to 2000 June.
We report the observations, including the first near-infrared
observations of this source yet published, 
in Section \ref{observations}, and the results and a discussion
are given in Section \ref{results}. 


\section{Observations} \label{observations}

All the observations took
place at European Southern Observatory (ESO), La Silla, Chile, 
except the NIR spectroscopy of 1999 September 17,
performed at UKIRT, Hawaii, and the optical imaging of
2000 June 24, performed at the 1.23 m telescope of the Centro
Astron\'omico Hispano Alem\'an at Calar Alto, 
with the CCD optical camera and exposure times
between 30 and 60s (more details on these observations
are reported in \citealt{marti:2001}).
The log of the optical and NIR imaging and spectroscopic observations 
is reported respectively in Tables \ref{table_observations_optique} and 
\ref{table_observations_infrarouge}.

The optical observations were performed 
with the 3.58m New Technology Telescope (NTT) 
equipped with the spectrograph and imaging camera EMMI RILD.
We imaged the source in V, R and I filters, and took
spectra with grism \#1 which gave a resolving power $R \sim 270$.
The exposure times were $\sim 5$ min with each filter for the imaging
and 15 min for the spectroscopy.
%
%
The ESO NIR observations 
were performed with the NTT, equipped with the 
infrared spectrograph and imaging camera Son of ISAAC (SOFI).
The imaging was taken through J, H and K$_{\rm s}$ filters, in combination
with the Large Field, giving a 4.9 x 4.9-arcmin$^2$ field of view, 
with a plate scale of 0.292 arc-seconds pixel$^{-1}$.
The spectra were taken with the Grism Red (GR) and $1\asec$ slit.
The exposure times were chosen as 9 min for the NIR imaging
(10 alternate images of 60s exposure time
each offset from the center by nearly 
$30\asec$ to the East, North, West, South,
following the standard procedure) and 15 min for the NIR spectroscopy.
The combined magnitudes
are the result of these 10 co-added and median filtered 60s exposures, 
with random offsets and direction between each exposure. 
The conditions were photometric for most of the observations,
the seeing being typically 0.8 arcsec,
and the airmass was always between 1.006 and 1.2.

The images were processed using IRAF reduction software. Each of the
images were corrected by a normalized dome-flat field, and 
the NIR images were sky-subtracted
by a sky image created from combining with a median filter 
a total of 10 consecutive images.
The data were then analysed using the IRAF reduction task ``apphot'', taking
different apertures depending on the photometric conditions of the night. 

Absolute photometry was performed using 2 standard stars from the
new system of faint near-infrared standard stars (\citealt{Persson:1998}):
No 9164 (HST P565-C) and 9178 (HST S808-C).
Each exposure of these standard stars is the average of 7x1.2s integration
time frames, and this is repeated 5 times by offseting the images of 1 arcmin
to the North-West, North-East, South-East and South-West from the
central position, and the final image is the 
co-add and median filter of those individual frames.
%

\begin{figure}
\centerline{\psfig{file=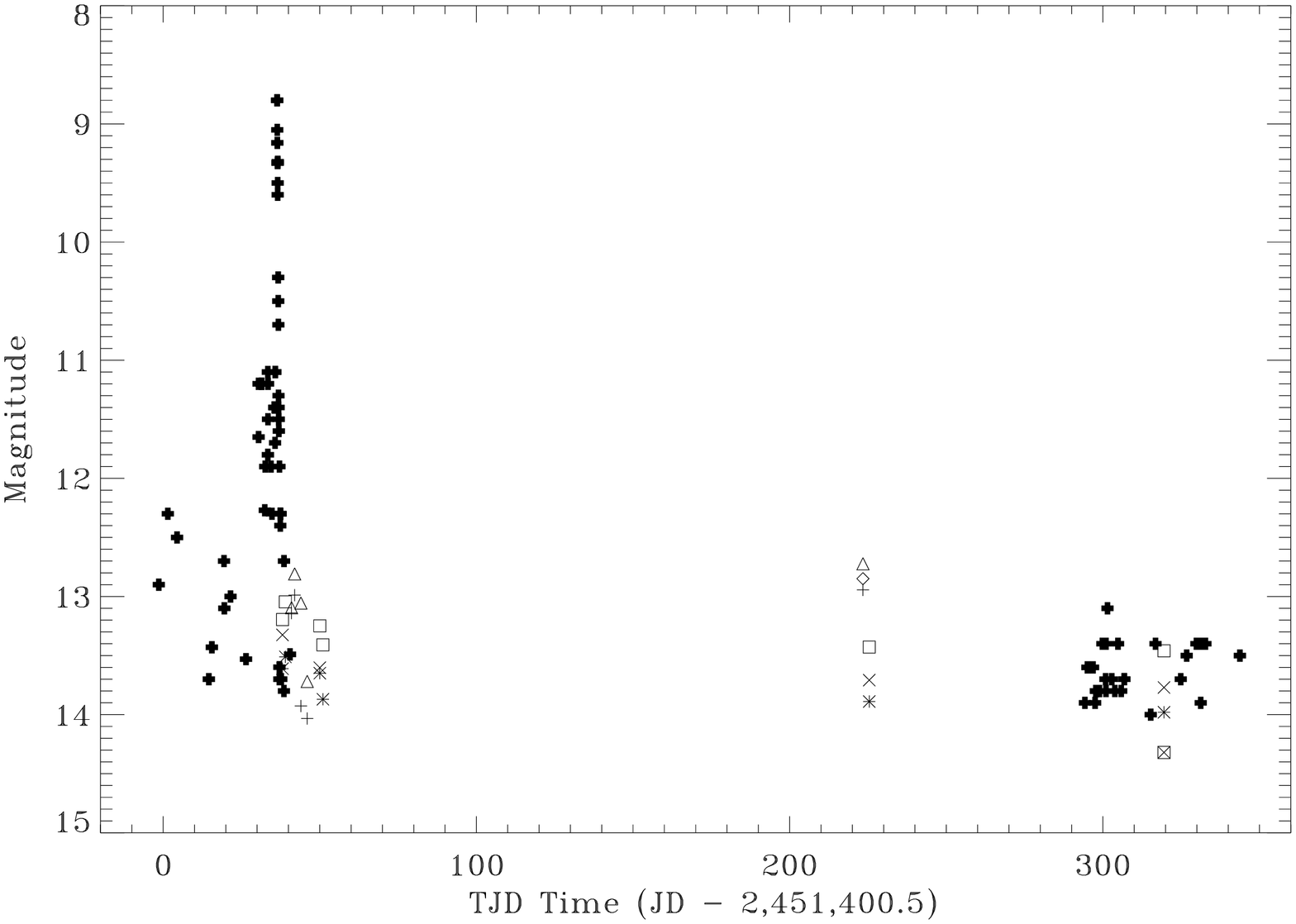,angle=+90.,width=9.cm}}
\caption[]{Optical and NIR observations of $\vqsqu$ 
from 1999 August to 2000 July. 
$\bullet$: VSNET, $\boxtimes$: B, $\ast$: V, $\times$:R, $\square$:I, $+$: J, $\diamond$: H, $\triangle$: K$_{\rm s}$ magnitudes. 
We first see a brief optical outburst which occured in 1999 August,
followed by the beginning of a modulation on 1999 Sept. 8 UT (= MJD 51429.5), 
which leads to the giant outburst of 1999 Sept. 15.7 UT (= MJD 51437).
The error bars of our observations are smaller than the size of the 
symbols used. TJD of 0 corresponds to 1999 August 10.
\label{optique_tout}}
\end{figure}

\begin{figure}
\centerline{\psfig{file=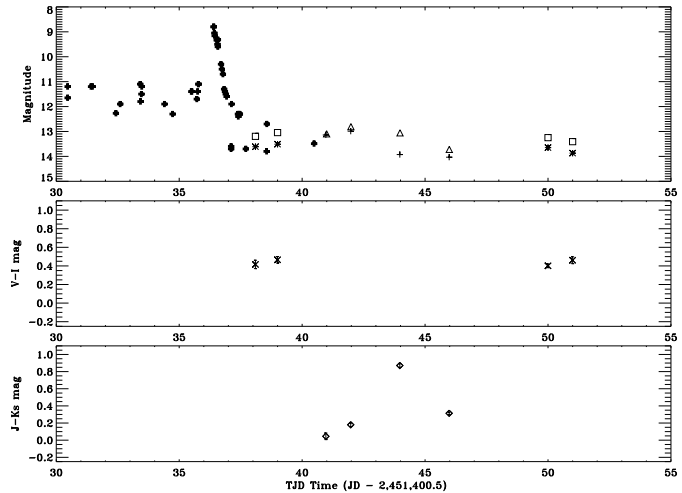,angle=+90.,width=9.cm}}
\caption[]{Top: Optical lightcurve of $\vqsqu$ during the outburst; 
Middle: V-I colour; Bottom: J-K$_{\rm s}$ colour.
$\bullet$:VSNET, $\ast$:V, $\square$:I, $+$:J, $\triangle$:K$_{\rm s}$,
$\times$: V-I, $\diamond$: J-K$_{\rm s}$ magnitudes. 
The error bars of our observations 
are smaller than the size of the symbols used.
TJD of 0 corresponds to 1999 August 10.
\label{figure_V-I}}
\end{figure}

\begin{figure}
\centerline{\psfig{file=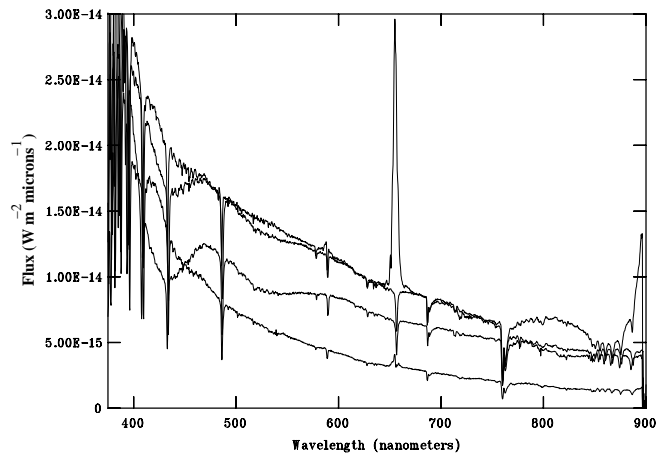,angle=-90.,width=11.cm}}
\caption[]{Flux calibrated optical spectra of $\vqsqu$ during the outburst.
From top to bottom the spectra were
 taken respectively on 1999 September 17.05, 18.0, 29.0 and 30.0 UT.
 \label{spec_opt_flux}}
\end{figure}

\begin{figure}
\centerline{\psfig{file=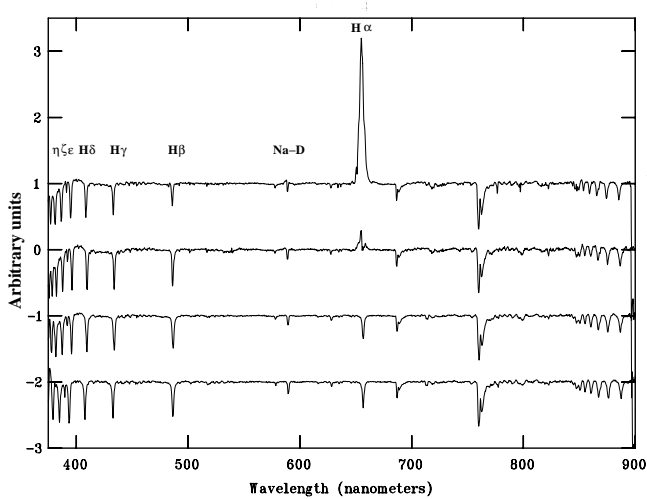,angle=-90.,width=12.cm}}
\caption[]{Normalized and offset optical spectra 
of $\vqsqu$ during the outburst. 
Same order as in Figure \ref{spec_opt_flux}.
The unlabelled absorption features are telluric ones.
\label{spec_opt_norm}}
\end{figure}

\begin{figure}
\centerline{\psfig{file=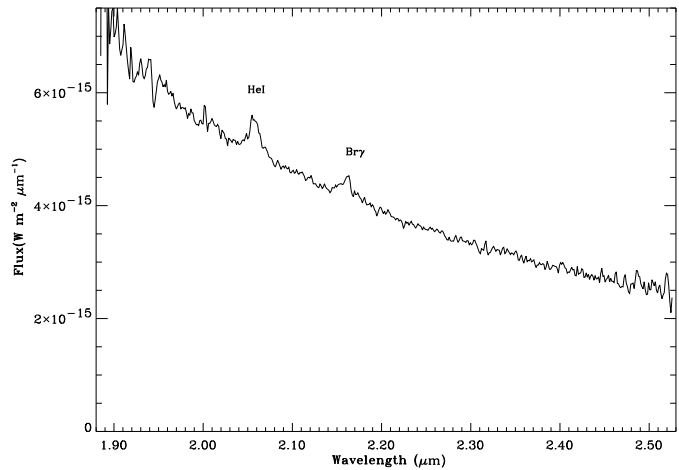,angle=90.,width=9.cm}}
\caption[]{Flux calibrated NIR spectrum of $\vqsqu$ during outburst. 
\label{spec_ir_charles}}
\end{figure}

The optical and NIR photometry is plotted in Figure \ref{optique_tout}.
An enlargement during the outburst interval (Sept. 19-24) and also the V-I 
and J-K$_{\rm s}$ colours are shown in Figure \ref{figure_V-I}.
The flux calibrated optical spectra are shown in 
Figure \ref{spec_opt_flux}, and again in normalized form 
in Figure \ref{spec_opt_norm}.
The flux calibrated NIR spectrum is shown in Figure \ref{spec_ir_charles}.
The strengths of the spectral features 
are given in Table \ref{spec_opt_lines}.

\begin{table*}
\begin{flushleft}
\begin{tabular}{|c|c|c|c|c|c|c|} \hline
{ Date}      & { MJD}     & { Inst}    & 
{ B}   & { V} & { R} & { I} \\ \hline 


16/09/99  & 51438.1 & EMMI      
& - & $13.60\pm0.10$ & $13.30\pm0.10$ & $13.20\pm0.09$ \\ 
17/09/99  & 51439.0& EMMI      
& - & $13.51\pm0.06$  &       -        & $13.04\pm0.08$  \\ 
%
%
28/09/99  & 51450.0& EMMI             
& - & $13.65\pm0.06$   & $13.60\pm0.01$ & $13.25\pm0.02$ \\ 
29/09/99  & 51451.0& EMMI 
& - & $13.87\pm0.06$  & -                & $13.41\pm0.08$  \\ 
21/03/00  & 51625.4 & EMMI 
& - & $13.89\pm0.11$ & $13.71\pm0.08$ & $13.43\pm0.04$ \\ 
24/06/00 & 51719.5 & C.A.
& $14.32\pm0.05$ & $13.98\pm0.05$   & $13.77\pm0.05$   & $13.46\pm0.05$ \\ 

\hline
\end{tabular}
\end{flushleft}
\caption[]{\label{table_observations_optique} Log of the 
optical observations.
}
\end{table*}

\begin{table*}
\begin{flushleft}
\begin{tabular}{|c|c|c|c|c|c|c|} \hline
{ Date} & { MJD}& { Inst} & { J     } &  { H} &  { K$_{\rm s}$} & { J-K$_{\rm s}$}     \\ \hline


17/09/99  & 51438.2& CGS4
& -                & -              & $12.5\pm0.3$     & - \\
19/09/99  & 51441.0& SOFI      
& $13.14\pm0.11$  & -              & $13.10\pm0.04$  & 0.04 $\pm$ 0.15 \\
20/09/99  & 51441.98& SOFI 
& $12.99\pm0.03$ & -              & $12.81\pm0.06$ & 0.18  $\pm$0.08 \\
22/09/99  & 51443.97& SOFI      
& $13.93\pm0.03$    & -              & $13.06\pm0.06$ & 0.87 $\pm$ 0.09 \\
24/09/99  & 51445.98& SOFI           
& $14.03\pm0.03$   & -              & $13.72\pm0.03$ & 0.31 $\pm$0.06 \\
20/03/00  & 51623.42& SOFI             
& $12.94\pm0.01$   & $12.849\pm0.01$  & $12.72\pm0.01$ & 0.22 $\pm$0.02 \\
%
\hline
\end{tabular}
\end{flushleft}
\caption[]{\label{table_observations_infrarouge} Log of the NIR observations.
The K$_{\rm s}$ magnitude of Sept. 17 has been estimated from the CGS4 spectrum
shown in Figure \ref{spec_ir_charles}
\citep{charles:1999a}.
}
\end{table*}

\begin{table}
\begin{center}
\caption[]{\label{spec_opt_lines} Optical emission and absorption 
line parameters observed during the outburst of $\vqsqu$. 
The flux is in $\WMMUM$. The typical uncertainties are 
$\sim 0.05 \rm \AA$ for Obs. $\lambda$ and 
$\sim 0.1 \rm \AA$ for EW and FWHM.}
\begin{tabular}{cccccc}
\hline
Date & Line & Obs. $\lambda$ & Flux      &  EW     &  FWHM       \\
1999 Sep. & Id. ($\rm \AA$) & ($\rm \AA$) & & ($\rm \AA$) & ($\rm \AA$ ) \\
\hline
17.05UT & H$\alpha$   6562 & 6565.23 &  9.0e-13 & -99.8 & 45.5    \\
        & H$\beta$    4861 & 4861.13 & -7.4e-14 &  4.6  & 13.1    \\
        & H$\gamma$   4340 & 4340.13 & -1.3e-13 &  6.8  & 14.1    \\
        & H$\delta$   4101 & 4101.83 & -1.6e-13 &  7.6  & 14.4    \\
        & H$\epsilon$ 3970 & 3970.16 & -2.1e-13 &  8.1  & 14.2    \\
        & H$\zeta$    3889 & 3889.63 & -2.2e-13 &  8.0  & 14.3    \\
        & H$\eta$     3835 & 3835.78 & -2.0e-13 &  7.5  & 13.4    \\
        & H$\theta$   3797 & 3798.43 & -1.5e-13 &  6.2  & 11.1    \\
        & H$\iota$    3770 & 3770.67 & -1.2e-13 &  5.6  & 11.5    \\
        & H$\kappa$   3750 & 3750.51 & -8.0e-14 &  4.2  &  9.1    \\
        & Na-D        5780 & 5785.60 & -5.6e-15 & $0.5 $&  8.9    \\
        & Na-D        5890 & 5898.32 & -2.2e-14 & $1.8 $& 10.1    \\
\hline
18.0UT  & H$\alpha$   6562 & 6544.13 &  2.1e-14 &  -6.9 & 30.2    \\
        & H$\beta$    4861 & 4861.08 & -7.4e-14 &  9.5  & 16.5    \\
        & H$\gamma$   4340 & 4340.82 & -1.3e-13 &  9.2  & 15.0    \\
        & H$\delta$   4101 & 4102.20 & -1.5e-13 &  8.7  & 13.8    \\
        & H$\epsilon$ 3970 & 3970.20 & -1.9e-13 &  9.4  & 14.1    \\
        & H$\zeta$    3889 & 3889.35 & -2.0e-13 &  9.0  & 14.4    \\
        & H$\eta$     3835 & 3835.29 & -1.7e-13 &  7.7  & 11.9    \\
        & H$\theta$   3797 & 3797.50 & -1.5e-13 &  6.8  & 10.8    \\
        & H$\iota$    3770 & 3770.00 & -1.4e-13 &  6.4  & 10.6    \\
        & H$\kappa$   3750 & 3749.41 & -1.0e-13 &  5.6  &  8.6    \\
        & Na-D        5780 & 5778.29 & -1.8e-15 &  0.4 &  9.0    \\
        & Na-D        5890 & 5891.15 & -8.6e-15 &  2.0  & 11.4    \\
\hline
29.0UT  & H$\alpha$   6562 & 6561.86 & -6.4e-14 &  7.0  & 20.1    \\
        & H$\beta$    4861 & 4861.12 & -1.5e-13 &  9.4  & 19.2    \\
        & H$\gamma$   4340 & 4339.74 & -1.6e-13 &  9.5  & 17.8    \\
        & H$\delta$   4101 & 4101.30 & -2.1e-13 &  9.9  & 17.6    \\
        & H$\epsilon$ 3970 & 3970.08 & -2.4e-13 &  9.4  & 15.6    \\
        & H$\zeta$    3889 & 3889.70 & -2.2e-13 &  8.7  & 14.9     \\
        & H$\eta$     3835 & 3835.86 & -1.8e-13 &  8.6  & 14.3    \\
        & H$\theta$   3797 & 3799.13 & -1.1e-13 &  6.5  & 11.4    \\
        & H$\iota$    3770 & 3772.49 & -5.9e-14 &  4.8  & 10.7    \\
        & H$\kappa$   3750 & 3751.25 & -3.9e-14 &  4.0  &  9.8    \\
        & Na-D        5780 & 5779.41 & -6.4e-15 &  0.5  & 11.6    \\
        & Na-D        5890 & 5891.97 & -2.3e-14 &  2.0  & 12.8    \\
\hline
30.0UT  & H$\alpha$   6562 & 6562.02 & -5.0e-14 &  7.4 & 19.3    \\
        & H$\beta$    4861 & 4860.46 & -1.0e-13 & 10.1  & 19.9    \\
        & H$\gamma$   4340 & 4340.10 & -1.1e-13 &  9.3  & 16.8    \\
        & H$\delta$   4101 & 4101.53 & -1.1e-13 &  9.3  & 16.1    \\
        & H$\epsilon$ 3970 & 3969.79 & -1.3e-13 &  9.4  & 14.9    \\
        & H$\zeta$    3889 & 3889.43 & -1.3e-13 &  8.6  & 14.6     \\
        & H$\eta$     3835 & 3835.40 & -1.2e-13 &  8.6  & 14.2    \\
        & H$\theta$   3797 & 3798.30 & -7.9e-14 &  6.3  & 11.2    \\
        & H$\iota$    3770 & 3770.75 & -5.6e-14 &  5.2  & 10.3    \\
        & H$\kappa$   3750 & 3749.82 & -4.4e-14 &  4.8  & 11.0    \\
        & Na-D        5780 & 5777.52 & -4.0e-15 & 0.5   &  9.1    \\
        & Na-D        5890 & 5890.62 & -2.1e-14 & 2.7   & 14.1    \\
\hline
\end{tabular}
\end{center}
\end{table}



\section{Results and discussion} \label{results}


After the big outburst (from V=14 to 8.8), 
there was still some flaring activity in V, R and I 
with variations of $\sim 0.5$ mag but no significant change in the colours.
In the NIR there was also 
some flaring activity with variation of $\sim 1$ mag in J and K$_{\rm s}$,
and a significant change in J-K$_{\rm s}$ during the post-outburst
interval
(between 5 and 8 days after the giant burst, see Fig. \ref{figure_V-I}).
This change is due to a decrease in J band much faster
than in K$_{\rm s}$, and during more than 
3 days, the source emits mainly in K$_{\rm s}$.
This can be explained either by the emission of a jet,
of the appearance of heated dust, or even by the interaction 
with the interstellar medium, as we will discuss later. 
We had observed the same phenomenon in $\grs$, which we had interpreted
as evidence for an extended cocoon of dust around the source 
\citep{mirabel:1996a}.
Later observations by Chandra \citep{lee:2002}
and ISO \citep{fuchs:2001} confirmed this presence of dust.

The first striking fact is that on a timescale of one day,
the optical lines were changing from emission to absorption.
All the H lines from the Balmer series from H$\alpha$ to H$\kappa$ 
are detected.
The H$\alpha$ emission line is extraordinarily strong and broad: 
one day after the outburst, its
equivalent width was $\sim 100 \rm \AA$, corresponding to a FWZI
(Full Width at Zero Intensity) of $4560 \kms$ with a blue wing.
This broad H$\alpha$ emission line
suggests the presence of a high velocity outflow component.
The single peaked H$\alpha$ profile suggests a low
inclination angle. 

There was also a weak He I $5876 \rm \AA$ line.
The Na-D absorption line equivalent width of $0.45 \rm \AA$ gives 
E(B-V) = 0.25 (following \citealt{munari:1997}) implying 
$\nh = 0.14 \times 10^{22} \cmmoinsdeux$ (following \citealt{bohlin:1978}).
This value of E(B-V) is consistent with that derived by 
\cite{wagner:1999}.
The second interesting fact is that 
there is a strong variability of the lines, as already pointed out
by \cite{garcia:1999}, particularly in H$\alpha$, H$\beta$ and also He I.
The He II $4680 \rm \AA$ line was claimed to be prominent in emission 
closer to the outburst time 
by \cite{ayani:1999}. Since we could not detect it, this line
was also very variable.
We can also note the blue continuum, visible on the flux calibrated spectra,
suggesting a contribution from an accretion disk, or from a corona.

In the NIR, the He I ($2.06 \microns$) and Br$\gamma$ ($2.17 \microns$) lines 
were observed as broad emission lines by \cite{charles:1999a}, in a K-band
(1.9-2.5-micron) low-resolution (2.5-nm) UKIRT (+ CGS4) 
spectrum taken on 1999 Sept. 17.22 UT 
and shown in Figure \ref{spec_ir_charles}.
He I exhibited an equivalent width of $2.1$ nm, and Br $\gamma$ of 1.4 nm,
characteristic of LMXBs and with widths of
$\mbox{FWZI} = 5900 \kms$. 
The Br-$\gamma$ profile shows a clearly extended blue wing, again
suggesting a high velocity outflow component.
The continuum of our NIR spectrum taken on 1999 Sept. 19 was also blue.
However, our NIR spectrum shows only 
very faint HeI (equivalent width $6 \rm \AA$), He II and Br$\gamma$ (equivalent
width $1 \rm \AA$) lines,
which therefore appear to be also strongly variable.

The broad, strong emission lines and their variability suggest the
presence of a high-velocity outflow component ($\sim 6000 \kms$)
blown off from the accretion of matter onto the compact
object (see for example \citealt{shulz:2002}).
It is likely that this outflow component forms thereafter 
an expanding plasma shell or even a cocoon which could have produced 
the change we observed in the J-K$_{\rm s}$ colour.

        \subsection{Nature of the system: the companion star}

\begin{figure}
\centerline{\psfig{file=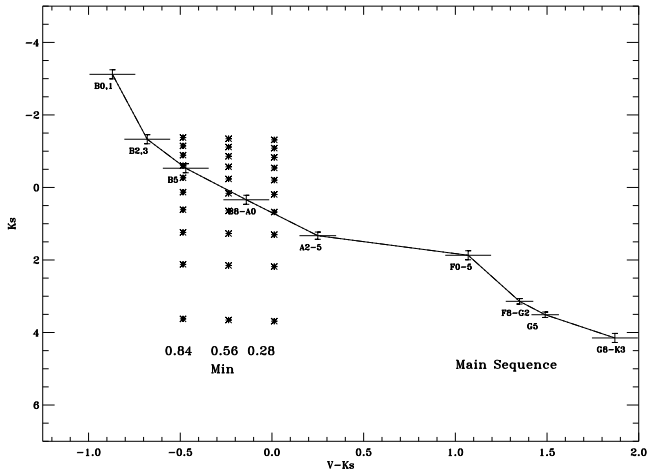,angle=+90.,width=9.cm}}
\caption[]{Colour-magnitude [V-K$_{\rm s}$,K$_{\rm s}$] diagram. *: Minimum absolute magnitudes 
of $\vqsqu$. 
+: typical main sequence stars \citep{ruelas-mayorga:1991}. 
0.28, 0.56 and 0.84 are the visual absorptions
corresponding respectively to column densities of
$0.05$, $0.1$ and $0.15 \times 10^{22} \cmmoinsdeux$.
From bottom to top the asterisks correspond respectively to a distance
of the source increasing from 1 to $10 \kpc$.
This shows that the distance is constrained to $3 < d < 8 \kpc$,
and that the spectral type is in this case
consistent with an early type B3 - A2 main sequence star.
\label{hr}}
\end{figure}

We plot on a colour-magnitude diagram (CMD, Figure \ref{hr})
the optical and NIR absolute magnitudes when the source was faint.
In this figure the absolute magnitudes are computed with
three different values of absorption
and ten different values of distance from $1-10 \kpc$.
The values of the absorption cover the range
$\nh = 0.05$ to $0.15 \times 10^{22} \cmmoinsdeux$,
derived by combining our observations and {\it Beppo-SAX} results 
\citep{in'tzand:2000a}.
If we constrain the companion 
to be a main sequence star, 
its location on the CMD, 
taking into account the uncertainty in the absorption,
suggests that the distance is constrained to
$3 < d < 8 \kpc$.
Its spectral type is in this case
consistent with an early type B3 - A2 V main sequence star.

However, it is interesting to note that the companion star could
be crossing the Hertzsprung gap, given the similarities of the system
with $\gro$ (\citealt{chaty:2002b}; \citealt{kolb:1997} 
and U. Kolb, private communication), 
although the mass of the companion star in the case of $\vqsqu$
is larger than for GRO J1655-40.
In this view, the location on the CMD would be above
the main sequence, the distance of the object could
then be larger than $3 < d < 8 \kpc$, this range
becoming its minimum distance, and the spectral type would be
B3 - A2 IV.
In both possibilities the mass is constrained between 2 $<$ M $<$ 10 $\Msol$,
suggesting that it is an IMXB or a HMXB.
This is consistent with the A2 V type at $6.1 \kpc$ 
derived by \cite{orosz:2001a}.

           \subsection{Interaction with surroundings}

If the elongation seen in the radio was a moving component, 
the proper motion was between $224 < \mu < 788$ mas/d depending on the
exact time of the ejection.
For the sake of discussion, we will take the lower limit, assuming
that the approaching (brighter) condensation exhibits
$\mu_a = 224$ mas/d.
Since 
from our results $D \geq 3 \kpc$, we conclude that the apparent
velocity in the plane of the sky would be strongly superluminal, with
$v_a$ greater than $4c$. 
However, 
no movement of this elongation was detected between Sept. 16.02 and 24.1.
This suggests an interaction between matter
ejected before the X-ray outburst and its surroundings at 
$0.25 \asec$ $\geq 1.5 \times 10^3$ AU from the source
at the distance of 6 kpc.
This is possible if the ejections began to take place at least 10 days before
the radio detection e.g. on September 8, and we can
see from Figure \ref{figure_V-I} that the source was already 
showing some activity in the optical at this date.
This interaction with the surroundings is supported by the increase in
J-K$_{\rm s}$ between 5 and 8 days after the outburst.

It then seems that the source activity was not as sporadic
as it appears at first glance.
Indeed, a previous optical outburst 
occured in 1999 August \citep{watanabe:1999}, 
and {\it RXTE} detected this source $\sim 270$ days
before the giant outburst \citep{in'tzand:2000a}.
Furthermore, at least 5 days before the giant outburst, 
$\vqsqu$ was, in the optical band, continuously
2 magnitudes brighter than immediately after, showing a modulation
at the orbital period, and with no X-ray emission
(typical upper limit of 12 mCrab, \citealt{in'tzand:2000a}).
All these facts, combined with the high X-ray variability,
show that, even if the \citet{orosz:2001a} results suggest that the accretion
in the system is of Roche-lobe overflow type, 
there is also the possibility that mass transfer in this
source is occuring through irregular wind accretion from the companion star.
This would not be too surprising in the case of an IMXRB/HMXRB system.

At a distance of $6 \kpc$, the maximum
luminosity of the source is $\sim 4 \times 10^{38} \ergs$, which is
close to the Eddington limit of a $\sim 10 \Msol$ object 
($1.3 \times 10^{39} \ergs$). 
If the mass transfer rate is highly super-Eddington
such a wind could arise.
This wind could be the reason why surrounding matter
was present, allowing the interaction between further ejections
and surrounding matter to take place.
In this case, the companion star is more likely to be a main sequence
star (U. Kolb, private communication).

Finally, it is interesting to note that \citet{marti:2001} observed
$\vqsqu$ in order to look for minute to hour variability, 
discovering 0.05 mag variability on a timescale of hours.
Among the different interpretations considered, 
they suggested that this variability could originate in an extended corona
surrounding the jets, by analogy with SS 433 \citep{zwitter:1991}.
We mention the possibility that the dust forming this corona 
could have its origin directly in the jets of X-ray binaries,
as in supernova ejecta.
If this interaction between the ejections 
and the surrounding medium is confirmed, 
this source could therefore be added
to the short list of microquasars where such an interaction has
been detected (see e.g. \citealt{chaty:2001a} and \citealt{mirabel:1999}).

            \subsection{A ``microblazar''?}

However, $\vqsqu$ seems different with respect to other microquasars, 
since in the latter
the outbursts normally fade more slowly, often lasting for weeks.
The only similar outburst was that from $\cicam$,
with an e-fold decay time of $\sim 0.5$ days \citep{belloni:1999},
its companion star being a symbiotic B star with an irregular wind.
Therefore $\vqsqu$ resembles several sources in behaviour,
but differs from them in other aspects: one possibility to explain
this is that it might be a ``microblazar'', 
i.e. a microquasar whose jet is pointing
towards the observer (see e.g. \citealt{mirabel:1999}).
\cite{orosz:2001a} already mentioned this possibility, based
on the new determination of the distance
of $\vqsqu$. Indeed, assuming that the radio component was moving 
\citep{hjellming:2000} implies that the jets
would then exhibit apparent velocities of $\sim 10$ c.
These large apparent velocities are consistent with the rapid variability 
in radio reported during the 2002's outburst 
(M. Rupen, VSNET communication), implying 
either extremely large Lorentz factors or a jet coming from a microblazar.
If $\vqsqu$ were a microblazar,
we also point out that we would not expect to see any Doppler shifted
emission lines from the ejecta, which is consistent with our spectra 
(Fig. \ref{spec_opt_norm}), as these lines would have a much 
larger blue/redshift than high-inclination systems.
For instance, assuming a plausible jet velocity of $0.95c$
with $\leq 10 \deg$ angle to the line of sight 
(as expected in a microblazar),
the Doppler shifted wavelengths of $H\alpha$ would appear in the UV
($\sim 1000 \rm \AA$) and NIR ($\sim 4 \microns$) 
for the approaching and receding jet, respectively. 

\section{Conclusions}

We have observed the source $\vqsqu$ during its 1999 September
outburst at optical and NIR wavelengths,
deriving E(B-V) = 0.25 and $\nh = 0.14 \times 10^{22} \cmmoinsdeux$.
By plotting our optical and NIR colours we have constrained the distance to 
$3 < D < 8 \kpc$, 
and the companion star would then be a main sequence star of spectral type 
B3 - A2 V.
If the source is crossing the Hertzsprung gap,
this determination of the distance would become its minimum distance,
and the spectral type of the companion star would be B3 - A2 IV. 
The system is therefore an IMXB or a HMXB.
From the NIR colours, and the optical spectra,
there is a strong suggestion of interaction of the ejecta
of the source with its surroundings.
This surrounding matter could have originated from an
outflow created by fluctuations around the central object,
and in this case the companion star would more certainly
be a main sequence star.
Further observations would be useful to confirm the existence
of surrounding matter.
The bright X-ray outburst of $\vqsqu$ in 1999 could have remained
unnoticed, because of its very short duration, if the optical
detection had not occured, and if the source was not located so close.
This means that there must be many similar (BH?) objects in our Galaxy,
most of them unnoticed when in outburst, because of their short duration
and faintest flares.
This type of sources will be targets of prime importance 
for present and future high-energy missions, 
as for example the {\it INTEGRAL} telescope.

\section{acknowledgements}
SC thanks Rob Hynes for pointing out this new
flaring source on September 1999, Bob Hjellming
for all the spontaneous communications he gave on the radio
observations of this source, and Ulrich Kolb for many stimulating
discussions. 
SC is very grateful to the ESO staff and particularly to the NTT team
(Leonardo Vanzi, Olivier Hainaut, St\'ephane Brillant and Vanessa Doublier),
for their availability and skills to perform service observations 
for Target of Opportunity programs (63.H-0493 and 64.H-0382, 
PI S. Chaty).
PAC and TS thank Tom Geballe for his assistance in obtaining 
the UKIRT CGS4 spectrum.
The United Kingdom Infrared Telescope is operated by the 
Joint Astronomy Centre on behalf of the U.K. Particle Physics 
and Astronomy Research Council.
We thank VSNET for all their alerts on $\vqsqu$
and their optical data used in Figures \ref{optique_tout} and \ref{figure_V-I}.
SC and PAC gratefully acknowledge
support from grant F/00--180/A from the Leverhulme Trust.
JM acknowledges partial support by DGICYT (AYA2001-3092)
and by Junta de Andaluc\'{\i}a (Spain), he has also been aided in this work
by an Henri Chr\'etien International Research Grant administered by the
American Astronomical Society.
IFM acknowledges support from grant PIP 0049/98 and Fundacion Antorchas. 
%
%

\bibliographystyle{aa}
\bibliography{science}

\end{document}